# Unveiling the shared grey matter signature between Alzheimer's and Parkinson's Disease

Vishaak Gangasandra[1,2*], Finn Blain[1,2*], Elise Delzant[3], Michelle K Lupton[1,4,5], Miguel E. Renteria[1,2,4], Sarah Medland[1,6,7], Baptiste Couvy-Duchesne[1,3,4,8]

[1]: Brain & Mental Health Program, QIMR Berghofer Medical Research Institute, Brisbane, QLD, Australia

[2]: Medical School, Faculty of Health, Medicine and Behavioural Sciences, the University of Queensland, Brisbane, Australia

[3]: Sorbonne University, Paris Brain Institute, CNRS, Inria, Inserm, AP-HP, Hôpital de la Pitié-Salpêtrière, Paris, France

[4]: School of Biomedical Sciences, Faculty of Health, Queensland University of Technology, Brisbane, Australia

[5]: School of Biomedical Sciences, Faculty of Medicine, The University of Queensland, Brisbane, Queensland, Australia

[6]: School of Psychology, The University of Queensland, Brisbane, Queensland, Australia

[7]: School of Psychology and Counselling, Queensland University of Technology, Brisbane, Queensland, Australia

[8] : Institute for Molecular Bioscience, The University of Queensland, St Lucia, Queensland, Australia

* these authors contributed equally

**Corresponding author**: Baptiste Couvy-Duchesne, baptiste.duchesne@qimrb.edu.au

Address: QIMR Berghofer, 300 Herston Rd, 4006 Brisbane, Australia

## Abstract

INTRODUCTION. This study presents the first quantification of vertex-level grey-matter associations between Alzheimer's disease (AD) and Parkinson's disease (PD) using high-resolution brain maps aggregated from large MRI datasets. The aim is to identify shared neuroanatomical signatures between the two diseases.

METHODS. Leveraging a novel statistical framework (SumR2 regression), adapted from genetic correlation analysis, we estimated the shared neuroanatomical signature (grey-matter correlation: rGM) between AD and PD.

RESULTS. A significant positive brain-wide grey-matter correlation (rGM=0.24, 95%CI 0.20–0.28) was observed between AD and PD. This correlation was further observed across disease stages and replicated using UK Biobank data. We located 9 vertex-wise clusters (106 vertices) that contribute to the significant rGM, highlighting reduced thickness in the bilateral putamen and right accumbens as associated with both AD and PD.

DISCUSSION. Our findings suggest that shared neuroanatomical features emerge early in neurodegeneration and have implications for early screening, disease monitoring, and targeted interventions.

**Acknowledgements**

Individual-level data used in the preparation of this article were obtained on [2024-08-30] from the Parkinson's Progression Markers Initiative (PPMI) database (www.ppmi-info.org/access-data-specimens/download-data), RRID:SCR_006431. For up-to-date information on the study, visit www.ppmi-info.org. PPMI – a public-private partnership – is funded by the Michael J. Fox Foundation for Parkinson's Research and funding partners, including 4D Pharma, Abbvie, AcureX, Allergan, Amathus Therapeutics, Aligning Science Across Parkinson's, AskBio, Avid Radiopharmaceuticals, BIAL, BioArctic, Biogen, Biohaven, BioLegend, BlueRock Therapeutics, Bristol-Myers Squibb, Calico Labs, Capsida Biotherapeutics, Celgene, Cerevel Therapeutics, Coave Therapeutics, DaCapo Brainscience, Denali, Edmond J. Safra Foundation, Eli Lilly, Gain Therapeutics, GE HealthCare, Genentech, GSK, Golub Capital, Handl Therapeutics, Insitro, Jazz Pharmaceuticals, Johnson & Johnson Innovative Medicine, Lundbeck, Merck, Meso Scale Discovery, Mission Therapeutics, Neurocrine Biosciences, Neuron23, Neuropore, Pfizer, Piramal, Prevail Therapeutics, Roche, Sanofi, Servier, Sun Pharma Advanced Research Company, Takeda, Teva, UCB, Vanqua Bio, Verily, Voyager v. 25MAR2024 Therapeutics, the Weston Family Foundation and Yumanity Therapeutics.

Brain association maps for Alzheimer's disease were generated in a previous publication (DOI 10.1002/hbm.70089) that used data from ADNI, AIBL, ARWIBO, EPAD, MAS, MEMENTO, OATS, OASIS3 and PISA. Please refer to the article for full acknowledgements of these datasets.

Brain maps generated on the UK Biobank (publication doi: 10.1002/hbm.70372) used application 53185.

Subcortical brain maps were also provided by the ENIGMA-PD consortium (doi: 10.1038/s41531-024-00825-9). We thank all the ENIGMA-PD authors and contributors, and in particular Max A. Laansma, who shared the subcortical maps on behalf of his co-first authors Yuji Zhao and Eva M. van Heese, under the supervision of Ysbrand D. van der Werf and Boris A. Gutman. Please refer to both dataset articles for full acknowledgements.

SEM and BCD were supported by NHMRC grant 2025674.

Conflict of interest: the authors have nothing to disclose.

## Introduction

Although AD and PD are characterised by distinct patterns of protein accumulation, postmortem investigations have revealed the co-occurrence of tau pathology in PD[1] and Lewy body pathology in AD[2]. Several studies have further reported Aβ accumulation in post-mortem brains of patients with PD, mainly in those with Lewy Body Dementia (LBD) and to a lesser extent, Parkinson's Disease Dementia (PDD)[1,3]. Hepp et al. also suggested that Lewy body accumulation is positively correlated with Aβ, and consequently, one misfolded protein could propagate other pathophysiological protein-mediated degradation[1]. Other possible shared mechanisms between AD and PD include iron accumulation, oxidative stress, mitochondrial dysfunction (caused by $\alpha$-Synuclein and tau), and loss of function in locus coeruleus noradrenergic neurons[4].

In addition, contemporary research has identified several epidemiological and genetic risk factors shared between AD and PD. Beyond age (a major risk factor for PD and AD), shared risk factors include a history of traumatic brain injury, diabetes, physical inactivity, anxiety and depression[5,6]. Furthermore, the use of some medications (incl. psychoanaleptics, drugs for constipation, sex hormone modulators and vasoprotective medications) has been associated with increased risk of developing AD and PD[7].

A study of family history in AD and PD cases suggested possible shared genetics or familial environmental factors between the diseases[8]. Molecular genetic studies have identified the tau-encoding MAPT gene on chromosome 17 as a possible site of overlap between Parkinson's disease[9] and Alzheimer's disease[10,11] (although the link with AD remains debated, as MAPT seems more strongly linked to frontotemporal dementia with Parkinsonism or PDD[12,13]). Large-scale GWAS of AD and PD suggest a small genetic correlation between the diseases (rg = 0.13, SE = 0.092, $p = 0.15$, i.e. not significantly different from zero with the current sample size)[14]. However, an investigation of local genetic correlation identified 20 regions of the genome associated with both diseases[15], including the genes: SNCA, CLU, KAT8, and SETD1A, previously linked to familial or sporadic forms of the disorders. Other stratified analyses of GWAS results identified overlap in the human leukocyte antigen region and in variants within neuronal open chromatin regions[16]. These common risk factors and genetic predispositions suggest the potential for shared pathological processes and aetiology, leading to convergent neuroanatomical signatures. If the origins and progression of disease follow a similar path, then similar brain markers may exist for both diseases. We hypothesise that the effects of neurodegeneration may be detected by similar markers in grey matter at early disease stages (due to genetics and early risk factors), as well as later disease stages (as protein cascades may be linked).

Large-scale cross-sectional analyses of grey matter structure using structural MRI have revealed detailed cortical and subcortical Regions of Interest (ROI) associated with each disease[17–19]. Eighteen ROI measurements are associated with both AD and PD, including 13 in the same direction (i.e., smaller thickness or volume in cases than in controls; summarised in **SFigure 1**). A notable limitation of ROI-based analyses is that each atlas region is likely composed of several sub-regions (e.g., hippocampal subfields), which are involved in specific disease- and symptom-related pathways (e.g., episodic, semantic, or procedural memory). This means ROI-level analysis can suggest overlap between AD and PD even when the specific local biomarkers differ. Consequently, we sought to evaluate the overlap between AD and PD at a fine-grained vertex level.

We applied a novel statistical method developed by Delzant et al. (2025) to estimate grey-matter correlation (rGM) from vertex-wise association maps of cortical and subcortical measurements (thickness and surface area) associated with each disorder. This novel approach enables cross-trait, cross-cohort analyses, which would not be possible otherwise. Grey-matter correlation is a direct estimate of the shared neuroanatomical signature between the diseases, i.e., how much the same vertex-wise grey-matter measurements can account for the diseases. We estimated rGM across the disease stages to provide insights into the nature of the shared neuroanatomical signatures over time. We further sought to identify localised (vertex-wise) brain biomarkers associated with both diseases.

## Materials and Methods

### Ethics Approval

Under the National Statement on Ethical Conduct in Human Research and relevant University of Queensland policy (PPL 4.20.07), this project operates under a reuse of research data and has received a Notification of Exemption from Human Research Ethics Review (2023/HE001584).

Each dataset used in this article has been approved by the relevant ethics committee in the country of collection. Please refer to the online cohort descriptions for more information.

### Alzheimer's and Parkinson's disease neuroimaging samples

Our analyses focused on the brain signatures of cases (PD or AD) vs. controls, although we also considered, when available, different disease stages that may exhibit distinct neuroanatomical signatures.

Vertex-wise brain association maps for AD cohorts were sourced from Couvy-Duchesne et al. (2025)[17]. The manuscript included 11 cohorts of T1-weighted MRI (N=6,981 in the discovery cohorts used to generate brain maps, comprising AIBL (Australian Imaging, Biomarker and Lifestyle)[20], ARWIBO (Alzheimer's Disease Repository Without Borders)[21,22], EPAD (European Prevention of Alzheimer's Dementia)[23–25], MAS (Sydney Memory and Ageing Study)[26–28], OASIS3 (Open Access Series of Imaging Studies 3)[29] and OATS (Older Adults Twin Study)[30–32]). Data included 796 (11.40%) AD patients, 1,343 (19.24%) patients with mild cognitive impairment, and 4,663 (66.80%) healthy controls (HC). The global mean age was 69.18 (ranging from 60.6 to 72.5 years across the cohorts). The overall sex distribution was males N=3,040 (43.55%) and females N=3,941 (56.45%). In addition to AD dementia vs. cognitively normal (CN) controls, we included brain maps of mildly cognitive impaired (MCI) individuals, which were also compared to cognitively normal individuals. In addition, we used brain maps of the early AD stage. Early AD focuses on individuals who are CN or MCI individuals at imaging and compares those who remain stable to those who progress to AD dementia within 1, 2, 3, 4 or 5 years of the MRI (named progression to AD).

**Table 1: Sample Characteristics for the AD and PD vertex-wise datasets**

| Cohort names | Number of participants | Average age | Sex (Number female; % female) | Diagnosis status |
|---|---|---|---|---|
| **ADNI (1,2,Go,3) AIBL ARWIBO EPAD MAS OASIS3 OATS** | 6,981 | 69.2 | 3,941; 56.5% | 796 Alzheimer's Disease Dementia cases, 1,343 Mild Cognitively Impaired (MCI), 4,663 Cognitively Normal (CN). 419 MCI or CN progressed to AD within 5 years; 774 stable MCI/CN. **Brain maps generated in Couvy-Duchesne et al., 2025.** |
| **PPMI** | 2,459 | 64.1 | 1,088; 44.3% | 1,109 PD cases, 872 controls and 471 prodromal cases. **Brain maps generated in this manuscript.** |
| **UK Biobank** | 39,826 | 62.8 | 21,040; 52.8% | PD: 93 cases (Either self-reported or hospital-reported). 51% were diagnosed before the imaging visit (2.4 years before on average [range 25 years before - 6.0 years after]. AD: 32 hospital-reported cases (incl. 29 diagnosed after the imaging visit (2.6 years after on average and up to 5.8 years). **Brain Maps generated in Delzant et al., 2025.** |
| **ENIGMA-PD** | 3,851 | 62.4 | 1,505; 39.1% | 2,525 PD cases, 1,326 controls from 50 cohorts. **Brain maps (subcortical nuclei only) generated in Laansma et al, 2024.** |

We created vertex-wise brain association maps for PD using data from the Parkinson's Progression Markers Initiative (PPMI)[33]. We processed T1w MRI images using FreeSurfer 6.0 and the ENIGMA-shape package[34–36]. This resulted in 654,002 vertex-wise measurements per individual, including 299,881 cortical vertices ('fsaverage mesh' in FreeSurfer) with thickness and surface area information and 27,120 subcortical vertices each with radial thickness and surface area (see our previous publications that used this processing for more details and testing [17,37,38]). Our final PPMI sample (with complete T1w processing, **Table 1**) included N=2,459 participants with a mean age of 64.13 years (at baseline). The sample comprised 1,088 females (44.25%) and included 1,109 PD cases, 872 controls and 471 prodromal cases. We defined prodromal status based on the PRIMDIAG [primary diagnosis] field and included individuals who exhibited either motor (N=50), non-motor (N=8) symptoms, or synucleinopathy (N=358). Notably, most PD cases were in Stages 1 and 2 of the Hoehn & Yahr scale, and there were insufficient end-stage cases to warrant separate analysis.

We also accessed vertex-wise maps of the subcortical nuclei, generated by the ENIGMA-PD consortium that included data from 50 cohorts[39] (**Table 1**).

## Construction of Brain Maps in PPMI

Using PPMI data, we performed vertex-wise association mapping with OSCA[40] to test associations between each grey-matter measurement and disease status, controlling for covariates (age, sex, site/scanner, ICV, total cortical thickness, and surface area). To maximise power in PPMI, we generated brain association maps using 3,629 MRIs, as some individuals were imaged multiple times (97 with 4 visits, 184 with 3 visits, 511 with 2 visits, and 1,667 with a single visit). We fitted participant ID as a random effect to account for the fact that some individuals were imaged up to 4 times. Brain maps for the AD cohort were produced by Couvy-Duchesne (2025) utilising the same analysis software, processing and covariates [(29)].

## Comparison of Vertex-wise Association maps

We sought to compare the vertex-wise maps of AD and PD to identify shared localised brain biomarkers and shared neuroanatomical signature. To define statistical significance in the brain map, we used an optimal Bonferroni-corrected significance threshold (derived from permutations; $p=2.0*10^{-7}$)[41] . We focused on vertices significantly associated with PD and AD and described their brain locations using the Desikan-Killiany Atlas[42].

**Sum R2 Regression analyses**

We estimated grey-matter correlations (rGM) between AD and PD using a recent method developed by Delzant et al. (2025)[43], and implemented the brainMapR R package. Grey-matter correlation is a multi-trait extension of the morphometricity framework that can quantify the shared neuro-anatomical signature between traits[37]. Morphometicity quantifies the global association between traits/diseases and all grey-matter structures measurements. In contrast, grey-matter correlation (rGM) quantifies the shared morphometricity between two traits/diseases. In practice, rGM estimates the unbiased whole-brain correlation between effect sizes (between two traits/diseases), accounting for the morphometricity of traits (which biases the correlation estimate) and the correlation among vertices. The SumR2 regression method is adapted from LD-Score regression, a widely used method in genetics for estimating genetic correlation[44,45].

SumR$^2$ regression requires vertex-wise brain maps, and SumR$^2$, which can be calculated on reference cohorts (provided in BrainMapR). We used the average SumR$^2$ – the consensus SumR$^2$ across ADNI, AIBL, OASIS, and UK Biobank. We used existing vertex-wise brain maps from Couvy-Duchesne et al. (2025)[17] for AD and ad hoc-generated brain maps for PD to estimate rGM. In follow-up analyses, we used brain maps generated in the UK Biobank[41], as well as the subcortical maps provided by the ENIGMA-PD consortium[39].

## Results

### Grey-matter correlations reveal shared neuroanatomical signatures in AD and PD across disease stages.

First, we quantified the shared neuroanatomical signature between AD and PD across all brain-wide vertices. We reported the grey-matter correlation (rGM) estimated from brain association maps using the model developed by Delzant et al. (2025)[43], **Figure 1**. rGM quantifies the brain-wide correlation among brain association maps while accounting for the morphometricity of traits (which biases Pearson's correlation estimates[43,46]), and while also taking into account the correlation structure among vertex-wise measurements[43].

We identified a significant positive grey-matter correlation between AD and PD (rGM of 0.24, SE=0.02; 95% CI: 0.20 0.28 $p=6.5\times10^{-34}$; **Figure 1**), indicating a shared neuroanatomical signature in which some common vertices are associated with increased or decreased risk of both diseases. Significant rGMs were also observed across all stages of both AD and PD included in the analysis (**Figure 1**). We observed high rGM values between Prodromal PD and AD Conversion at 1 (rGM=0.79, 95%CI: 0.55 1.0), 2 (0.86, 95%CI: 0.64 1.0), 3 (0.99, 95%CI: 0.77 1.0), 4 (0.71, 95%CI: 0.45 0.95), and 5 (0.77, 95%CI 0.52 1.0) years [although larger SE~0.11, due to smaller sample sizes], suggesting that the shared neuroanatomical signature is present from early stages of the diseases. The rGM between continuous clinical scales,UPDRS Score for PD and MMSE for AD (rGM = -0.46), further confirmed the association we observed between disease status (cases vs. controls). The negative rGM is due to the MMSE scoring system (lower scores indicate greater cognitive impairment).

We sought to replicate the rGM associations using brain association maps derived from clinical records of AD and PD from the UK Biobank (n=38,643)[41,43]. We observed an rGM of 0.12 (SE=0.05, p=0.0078, 95%CI: 0.035-0.22) between PD from PPMI and AD from the UK Biobank, which is consistent with the rGM we estimated previously (**Figure 1**). However, unexpectedly, the rGM between AD (from clinical cohorts) and PD from the UK Biobank was negative: rGM = -0.20 (SE = 0.03, p-value = 4.3e-12, 95% CI: -0.26 -0.15). We attributed this result to the limited number of PD diagnoses in the UK Biobank (a few hundred cases) and to the fact that most were diagnosed prior to imaging, which likely indicates early-onset PD (likely monogenic) and represents different cases than those from PPMI. Indeed, the rGM between the PD maps (PPMI and UK biobank) was only 0.26 (SE=0.03, 95%CI: 0.20 0.31), which indicates that the PD diagnoses in the UKB do not represent the same construct as that from case control samples. In contrast, the rGM between the two AD maps (clinical cohorts and UK biobank, with age of onset >65, **Table 1**) was 0.64 (SE=0.03, 95%CI: 0.58 0.70), which suggests cases are more comparable, despite some known ascertainment in the UK biobank[47]. For completeness, we estimated the rGM using AD and PD brain maps generated from the UK Biobank, which also pointed towards a positive grey-matter association between AD and PD (rGM=0.59, SE=0.03, 95%CI: 0.52 0.66).

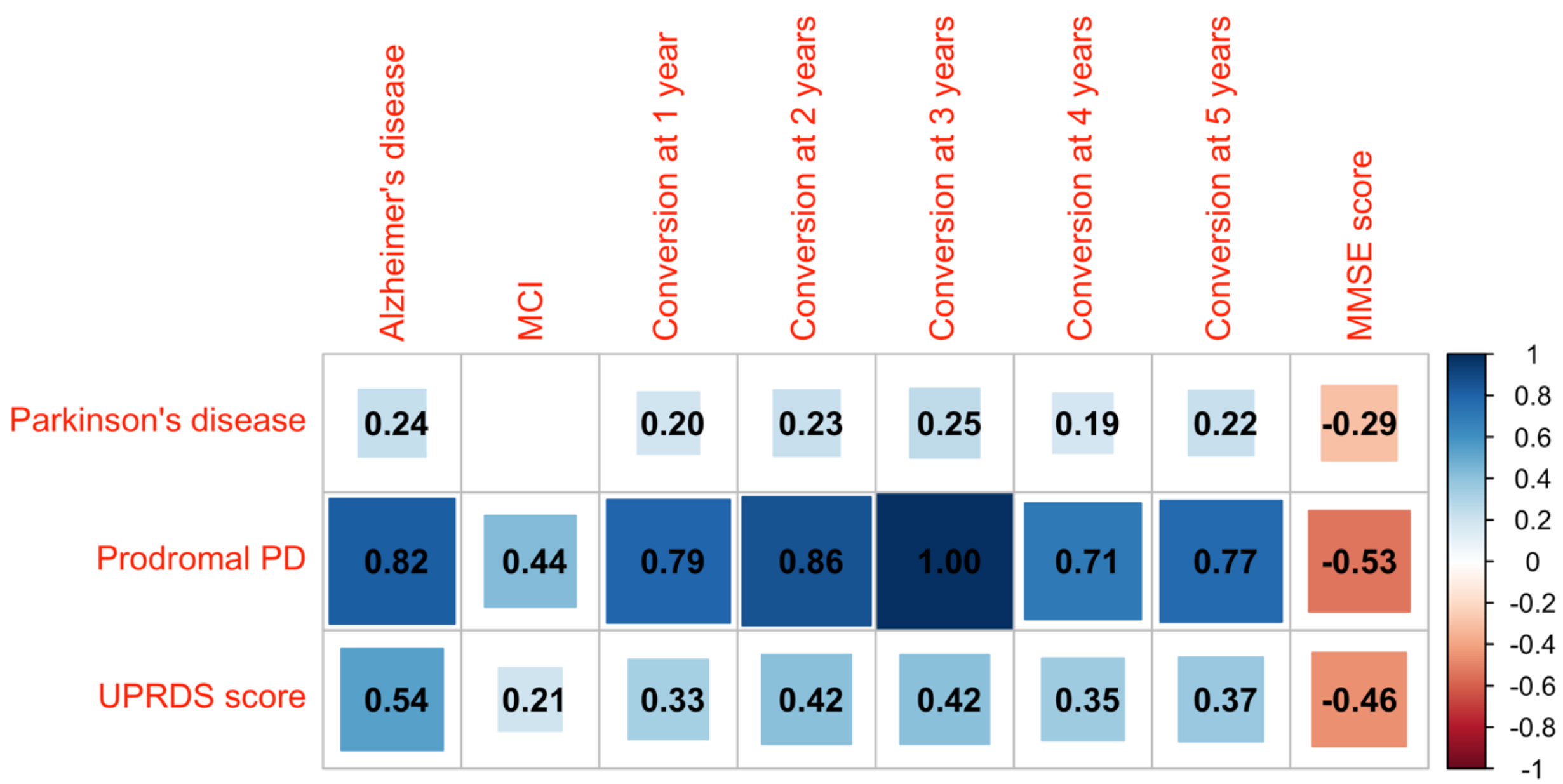

**Figure 1. Grey-matter correlations (rGM) between the different stages of AD and PD.** MCI: mild cognitive impairment. Conversion at 1-5 years corresponds to individuals imaged 1-5 years prior to a diagnosis of AD dementia, compared to individuals followed for the same amount of time but who did not progress. All rGM shown in colours are significant after multiple testing correction (pvalue <0.05/24). The negative rGM with MMSE reflects its coding: lower MMSE scores indicate lower cognition and greater disease severity.

**Vertex-wise associations identifies shared brain biomarkers in the putamen and nucleus accumbens thickness.**

Results from vertex-wise association testing revealed grey-matter structural differences between individuals with AD dementia and cognitively normal individuals. Using an optimal[41] Bonferroni significance threshold ($p<2*10^{-7}$), we identified 144 significant clusters associated with AD (comprising 7,822 significant vertices). These clusters were spread out across the cortex (mainly cortical thickness), including the entorhinal, fusiform, parahippocampal, precentral, paracentral, postcentral, superior temporal, middle temporal, infratemporal, cuneus, precuneus, posterior cingulate and isthmus cingulate gyri. In addition, clusters were detected in the thickness and/or surface area of the thalamus, putamen, pallidum, hippocampus, amygdala, and nucleus accumbens (**Figure 2, 3, STable 1**).

In contrast, vertex-wise association testing of PD versus controls in PPMI identified 13 significant clusters (64 significant vertices) corresponding to differences in grey-matter thickness in the pars triangularis, medial orbitofrontal cortex, precuneus, thalamus and pallidum, and in the left putamen (**Figure 2, 3, STable 2**). Vertex-wise associations from ENIGMA-PD[39] (**Table 1**), although limited to the vertices in the subcortical structures, identified additional associations with PD in left and right caudate thickness, right putamen and accumbens thickness, left thalamus area, and bilateral pallidum area (308 significant vertices; **SFigure 2**).

We identified 106 vertices associated with AD and PD (ENIGMA-PD), located in the left Putamen (33 vertices; 4 clusters), the right Putamen (70 vertices; 4 clusters), and the right Accumbens (3 vertices, 1 cluster). No vertex was significantly associated with both AD and PD when using PPMI only. All 106 significant vertices showed reduced thickness of these brain regions in AD and PD compared with controls (**Figure 4, STable 3**).

Next, we investigated brain regions where (non-overlapping) clusters of significant vertices were associated with each disease. We sought to explore whether the lack of overlap between significant vertices could be due to limited statistical power by visualising the correlation between effect sizes in those regions (**SFigure 3**). These regions included the subcortical thickness of bilateral thalamus, pallidum and of the left caudate, accumbens and the cortical thickness of the right cuneus (**SFigure 3**). In the right precuneus and left pallidum, we did not observe a strong correlation in effect sizes ($r=0.05$ and $r<0.12$), suggesting that different vertices are associated with AD and PD. Correlations between effect sizes were substantial in the bilateral thalamus thickness (r=0.32 and r=0.41), and consistent across samples (PPMI and ENIGMA-PD). This suggests that the vertices associated with one disease tend to exhibit effect sizes similar to those in the other disorder (**Figure 3**). Interestingly, the correlation between AD and PD effect sizes was inconsistent across PPMI and ENIGMA-PD in the right Pallidum, left Caudate and left Accumbens, which might be due to the use of different covariates (we controlled for average thickness in PPMI, in addition to ICV). Yet, for the right Pallidum thickness, both correlations with PPMI and ENIGMA-PD were negative, suggesting that vertices in this region are differently associated with the diseases (risk factors in AD but associated with reduced PD risk; **SFigure 3**).

Lastly, we sought to clarify whether ROIs associations in AD and PD (ROI average analyses; **SFigure 3)** are driven by the same vertices. Thus, we explored the correlation between AD and PD vertex-wise effect sizes in ROIs reported in Laansma et al. (PD)[18] and Couvy-Duchesne et al., (AD)[17]. For most ROIs, the correlation between effect sizes was small to negligible, which suggests that different vertices drive the associations detected in PD and AD (**SFigure 4**). Notable exceptions were putamen thickness (r=0.49 in the left, r=0.56 in the right; where we identified shared biomarkers), left putamen surface area (r=0.40), and left inferior temporal gyrus thickness (r=0.29), for which the correlations suggest that some vertices exhibit similarly large associations with both diseases.

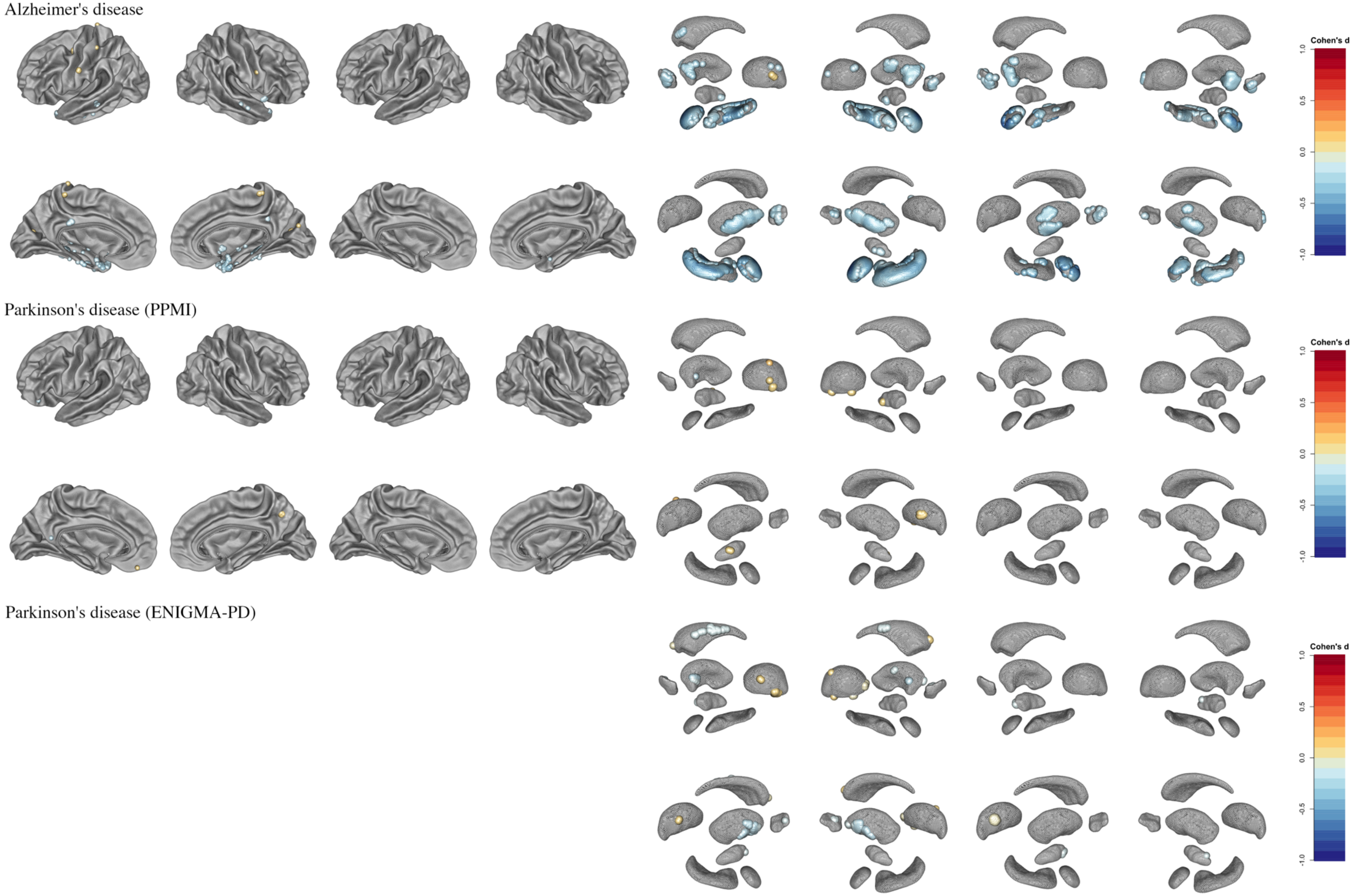

Alzheimer's disease
Parkinson's disease (PPMI)
Parkinson's disease (ENIGMA-PD)
Cohen's d
1.0
0.5
0.0
-0.5
-1.0
Cohen's d
1.0
0.5
0.0
-0.5
-1.0
Cohen's d
1.0
0.5
0.0
-0.5
-1.0

**Figure 2: Brain maps of significant vertex-wise associations with Alzheimer's Disease (dementia versus cognitively normal) (top figure) and Parkinson's disease (PPMI cases versus controls; middle figure, ENIGMA-PD results for subcortical vertices only – bottom figure).**
Cortical isosurface views are presented on the left (left hemisphere thickness, right hemisphere thickness, left surface area and right surface area, respectively), followed by subcortical isosurface on the right (left hemisphere thickness, right hemisphere thickness, left surface area and right surface area, respectively). Only the significant vertex-wise measurements after multiple testing correction ($p < 2x10^{-7}$) are shown. The association effect sizes correspond to Cohen's *d*, that is, the mean difference between groups in terms of standard deviation units of the vertex-wise measurement (0: Controls, 1: Alzheimer's/Parkinson's). In practice, we fitted (for AD and PD-PPMI) the model DiseaseStatus ~ a*covariates + b*vertexMeasurement_standardised, and converted the b into Cohen's d using the formula r=b/sd(DiseaseStatus) and d=2*r/sqrt(1-r**2), following the formulas in[48].

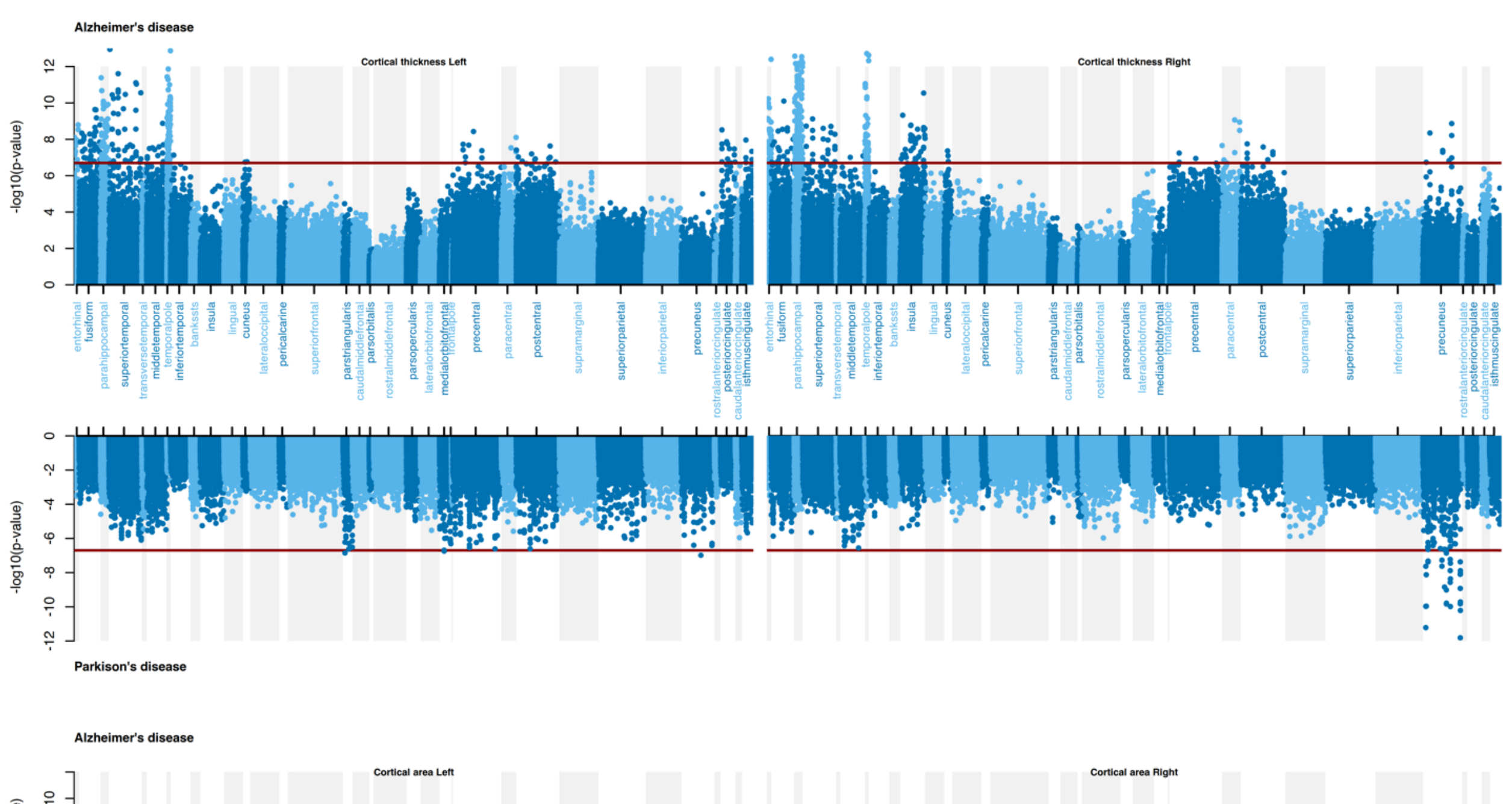

Alzheimer's disease
Cortical thickness Left
Cortical thickness Right
-log10(p-value)
Parkison's disease

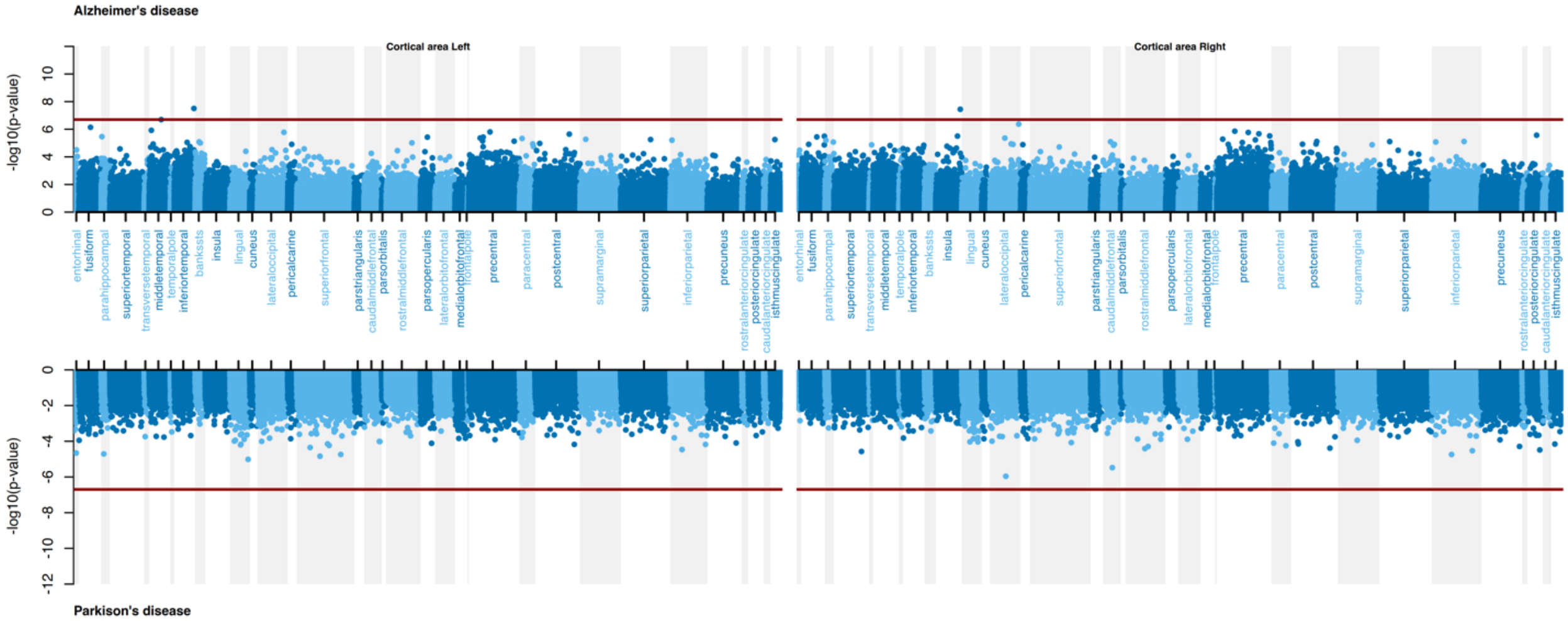

Alzheimer's disease
Cortical area Left
Cortical area Right
-log10(p-value)
Parkison's disease

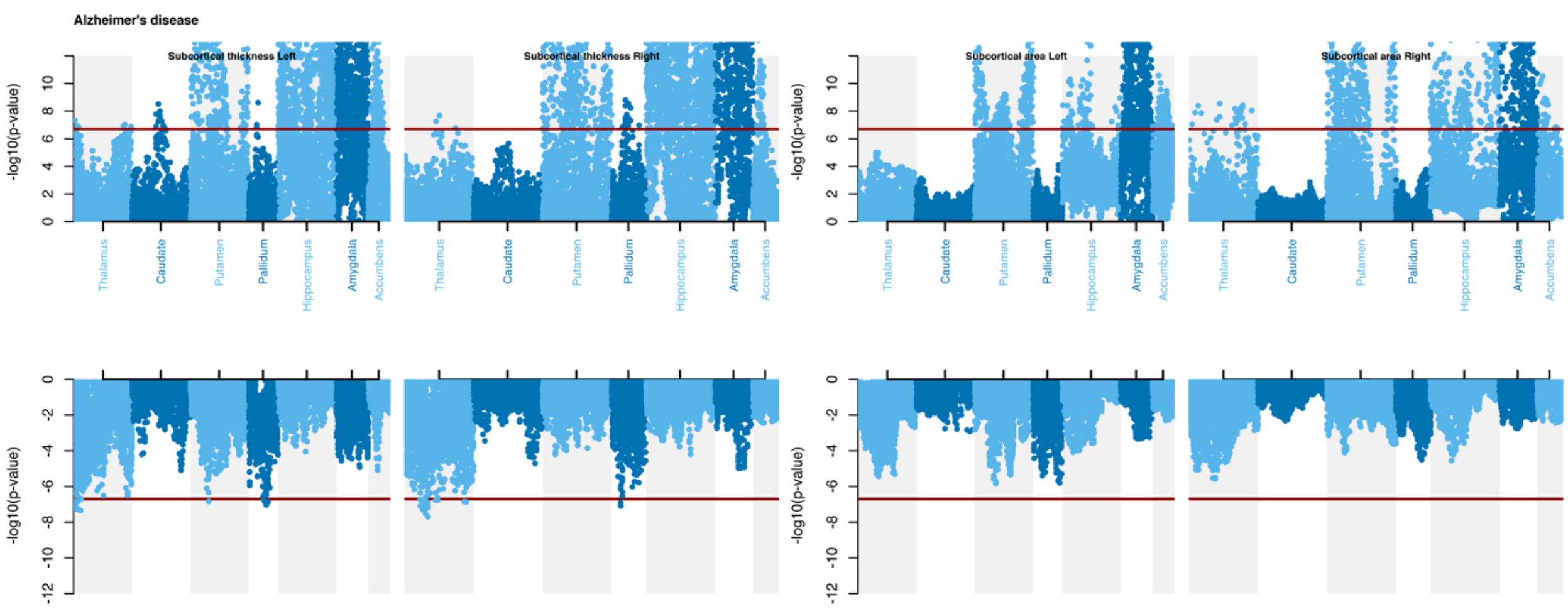

Alzheimer's disease
Subcortical thickness Left
Subcortical thickness Right
Subcortical area Left
Subcortical area Right
-log10(p-value)
Thalamus
Caudate
Putamen
Pallidum
Hippocampus
Amygdala
Accumbens
Parkison's disease

**Figure 3. Miami plot displaying vertex-wise association strength (-log10(pvalue)) in Alzheimer's Disease (top) vs. Parkinson's disease (PPMI; bottom).**

The red horizontal bars represent the significance threshold after multiple testing correction (pvalue<2.0E-7). Vertices crossing this bar indicate significant associations in the ROI. The different ROIs are highlighted by white/grey shading and the dark blue/light blue colours. From top to bottom, we show cortical thickness ROIs (left, then right), cortical surface area ROIs, and subcortical thickness and area.

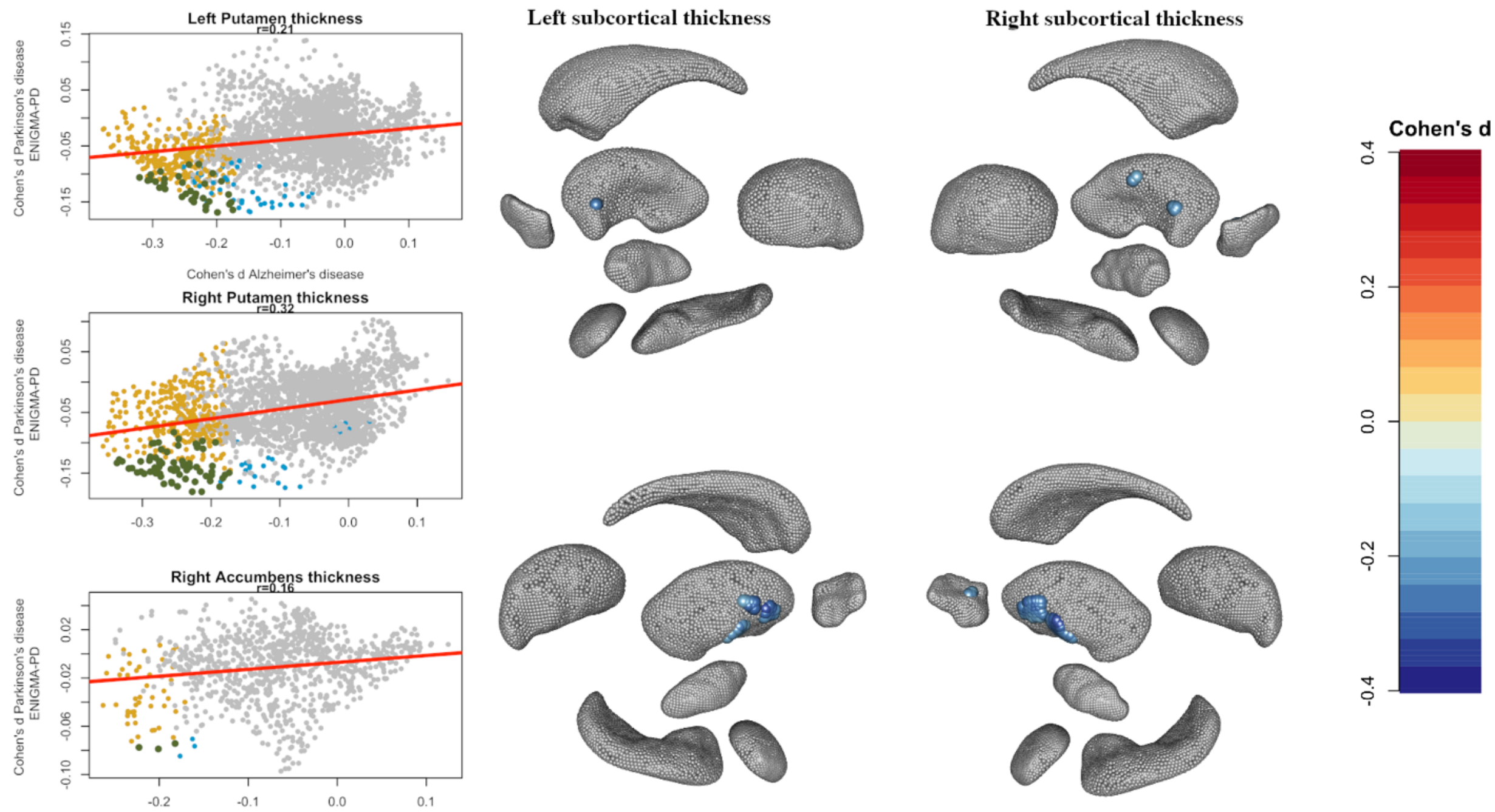

**Figure 4. Vertices associated with Alzheimer's and Parkinson's disease.**
Left panels show the correlation of Cohen's d within the ROIs of interest (left and right putamen thickness, right accumbens thickness). Significant vertices after multiple testing correction (p-value < 2.0E-7) are highlighted in yellow (vertices significantly associated with Alzheimer's disease) and blue (vertices significantly associated with Parkinson's disease in ENIGMA-PD). Vertices associated with both diseases are shown in green. Pearson's correlation between effect sizes is shown at the top of the plot, and the regression line is displayed in red.
The right panels show the location of those significant vertices (associated with AD and PD). Effect sizes for Alzheimer's are shown, but associations are in the same direction for PD. Outside view (top) and inside view (bottom) of the subcortical nuclei are shown.

## Discussion

We quantified, for the first time, the associations at the vertex level between grey-matter signatures of AD and PD. We applied a novel statistical method (SumR2 regression[43]) to estimate grey-matter correlation (rGM), which measures the brain-wide correlation between brain association maps while accounting for the complex correlation structure among vertex-wise measurements[43]. SumR2 regression leverages a statistical framework, also used in genetics to estimate genetic correlation[45], offering insight into the broader connectedness of diseases. As it relies on brain association maps rather than individual-level data, SumR2 regression is computationally efficient and enables cross-cohort, cross-disorder analyses without requiring information on disease comorbidities for a unique set of participants.

We found a significant positive brain-wide grey-matter correlation between AD and PD (rGM=0.24, 95% CI 0.20 0.28, **Figure 1**), suggesting that a fraction of the vertex-wise measurements are associated with increased or decreased risk for both disorders. We replicated this result using AD brain maps created from the UK Biobank (rGM=0.14, 95%CI 0.07 0.20). This result is noteworthy, as the maps generated from the UK Biobank rely on noisy clinical status, being limited by the age of the participants (median 67 years old at imaging, i.e., some are too young for accurate diagnosis of neurodegenerative disease) and imperfect linkage of clinical data from primary and secondary care sources[49]. The heterogeneity between clinical status from the UK Biobank and case-control datasets was confirmed by the within-disease rGM which was 0.64 (95%CI 0.58 0.70) for AD, but only 0.25 (95%CI 0.20 0.31) for PD. The heterogeneity in PD diagnoses in UK Biobank (compared to PPMI or ENIGMA-PD) is likely due to a fraction of cases being early-onset PD with a different neuroanatomical signature than late-onset cases.

We further identified grey-matter correlations between Prodromal Parkinson's Disease and early AD (up to 5 years prior to AD dementia diagnosis), suggesting that the shared neuroanatomical signature between the diseases is present from the early stages of neurodegeneration, further supported by a significant rGM between UPDRS and MMSE scores (rGM = -0.46, 95% CI -0.50 to -0.42; negative due to MMSE coding). We can hypothesise that the shared brain biomarkers in early disease stages may reflect some of the shared genetics[14,15] or known risk factors common to AD and PD[5,6]. Similarly, the shared brain markers of disease severity (MMSE and UPDRS) could reflect overlapping patterns of neurodegeneration driven by distinct biological processes involving amyloid-beta, tau and alpha-synuclein[1].

Progressing the interpretation and understanding of the shared neuroanatomical signature between AD and PD requires identifying the specific vertex-wise biomarkers shared between the diseases. Using subcortical maps from ENIGMA-PD, we identified 106 vertices across 9 clusters that showed reduced thickness in AD and PD. These clusters implicated the left and right putamen and the right accumbens (**Figure 4**).

Grey-matter structure of the putamen has been clearly identified as a brain biomarker of PD and AD, with large multi-cohort analyses reporting reduced volume in cases compared to controls[17,18], as well as vertex-wise reductions in putamen thickness[17,39]. Our results suggest that similar regions of the putamen may be affected in both diseases (**Figure 3, SFigure 3**), with thickness and surface area being smaller in cases than in controls (**Figure 2, 3**). Putamen structural alterations have been associated with motor symptoms in PD[39,50], and are thought to be caused by the loss of dopaminergic neurons in the substantia nigra pars compacta, which decreases dopamine input to the putamen[51]. However, the putamen's vertex-wise structure is also associated with cognition in PD patients, which aligns with the known functions of the putamen in emotion, motivation, and cognition beyond pure motor function. In Alzheimer's disease, the putamen structure has also been associated with the Functional Activities Questionnaire (FAQ) score[17], which measures individuals' ability to perform activities of daily living involving cognition and complex motor planning.

The nucleus accumbens (NAc) is an essential brain region of the ventral striatum responsible for dopaminergic-mediated reward, emotions and cognitive functions[53]. NAc atrophy in PD (sometimes referred to as Mavridis' atrophy[54]) has been well characterised as an early-stage biomarker influencing psychiatric and cognitive symptoms associated with the disease[54]. Similarly, the NAc structure has been directly correlated with AD clinical impairments (MMSE score)[55] , and with AD status[17], and reduced functional connectivity has been linked with deficits in cognitive function, in particular, episodic memory[56].

In addition, several cortical and subcortical regions were implicated in both diseases (showing significant vertex-wise associations, albeit from different vertices; **Figure 2, 3**). This suggests that better-powered vertex-wise association maps will likely detect shared brain markers of AD and PD in those regions. One such region, the thalamus, serves as the relay centre for the brain, integrating motor, sensory, and emotional signals from the body to the cerebral cortex[57,58]. In PD, the thalamus structure has been linked to cognitive and motor alterations, although different parts of the thalamus may be at play (anteroventral and ventral posterolateral nuclei linked with cognition; while motor impairments could be linked to lateralisation asymmetries in the centromedian, mediodorsal medial, pulvinar anterior, and pulvinar medial nuclei)[59]. In AD, the structure of the thalamus (in particular, the anteroventral nucleus, the ventral anterior nucleus, the mediodorsal nucleus, and the Pulvinar nucleus) has been associated with disease status, as well as with semantic and phonemic fluency and episodic memory[60]. Other possible brain regions that could contribute to the grey-matter correlation include part of the left inferior temporal gyrus thickness which is broadly implicated in semantic memory and visual processing. Larger neuroimaging studies are required to confirm this hypothesis and identify more localised (e.g., vertex-wise) grey-matter regions that contribute to the shared neuroanatomical signature of AD and PD. Interestingly, we also identified some brain regions (e.g., right pallidum thickness), where the negative correlation between vertex-wise effect sizes contradicts the positive rGM we estimated across the brain. This result suggests that grey-matter correlation may be driven by a subset of regions. Local rGM may be stronger in some regions (e.g., thalamus, putamen), absent in others (e.g., right precuneus thickness; **SFigure 3**), and negative in a few regions. More work is needed to extend our rGM framework to estimate local rGM.

Our results emphasise some of the limitations of ROI-based analyses of the brain, which lack fine-grained information and can be misleading in cross-disorder analyses aimed at identifying grey-matter biomarkers common to diseases. ROI-based analyses of AD and PD identified 18 regions whose average structure (thickness, surface area, or volume) was associated with both diseases[17,18]. Our vertex-wise analysis reveals that in 11 of those cases, the correlation between AD and PD effect sizes was minimal ($|r|<0.1$; **SFigure 3**), suggesting that the ROI level associations are driven by vastly different sets of vertices. In addition, ROI analysis can miss important regions where the effect is localised. For example, the average left thalamus volume was associated with PD but not significantly associated with AD in a meta-analysis of 10 cohorts, despite significant vertices being identified[17].

The impacts of our findings are wide-reaching across research and clinical translation. Our findings identify shared grey-matter pathways between the diseases and point towards subcortical areas (thalamus and putamen) for future investigation of disease similarities. These regions may have properties that attract misfolded proteins in neurodegeneration, or they may be independently susceptible to proteinaceous damage. In addition, our results raise questions about possible rGM with other neurodegenerative disorders, including vascular dementia, frontotemporal dementia or amyotrophic lateral sclerosis. Lastly, evidence of shared neuroanatomical signatures suggests that researchers could combine (e.g., meta-analyse) brain association maps from correlated disorders to increase the power to detect overlapping brain regions.

Clinically, the implications of AD and PD sharing similar early-stage progression are impactful. Identifying biomarkers shared between neurodegenerative diseases (especially if present in early disease stages) could aid early screening and tracking disease(s) progression, enabling better supportive strategies and potentially preventing or slowing progression. The identification of brain markers specific to each disease could also help clinicians better differentiate diseases early on, enabling patients to be started on definitive treatment years earlier.

Our study and interpretation of the results are limited by several factors. Firstly, the large datasets we used remain insufficient to identify many vertex-level markers and pinpoint all the specific grey-matter regions that contribute to the rGM. Estimating localised rGM could help advance the identification of shared brain regions[46], but this method is not yet available for cross-trait, cross-cohort analyses. The smaller sample sizes for prodromal PD or early AD, as well as late disease stages, further limit the identification of vertex-wise associated regions, which will eventually shed light on the early nature of the grey-matter overlap between AD and PD. In addition, our study relied on cross-sectional data and did not track subjects' brains across time. Finally, it is possible that some of the findings are due to individual differences unrelated to the disease (e.g. external confounders).

## Supplementaries

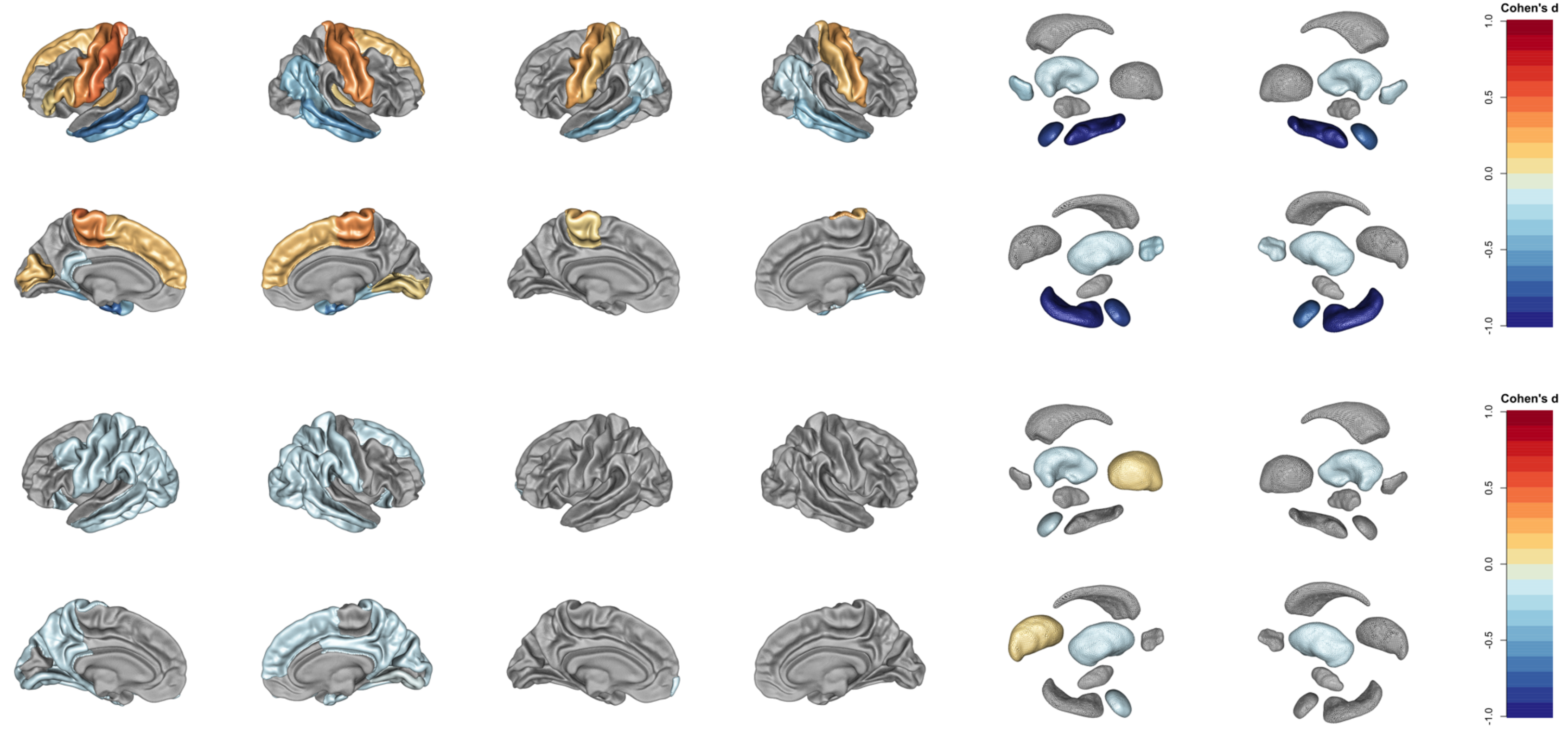

**SFigure 1: ROI associations between grey-matter structure and Alzheimer's disease (top) and Parkinson's disease (bottom).**

Associations have been reported in Couvy-Duchesne et al., 2025 (AD)[17] and Laansma et al. (2021)[18] for PD. Plots from left to right, display results for left cortical thickness, right cortical thickness, left and right surface area, left and right subcortical volumes. External (top row) and internal views (bottom row) are provided for each disease. For AD, we converted the associations reported (b - from the model ADcvHC ~ covariates + ROI_std) into cohen's d using the formula r=b/sd(AdvsHC) and d=2*r/sqrt(1-r**2)[48]. Eighteen ROI measurements were associated with both AD and PD, including 13 in the same direction (i.e., smaller thickness or volume in cases compared to controls). These 13 common biomarkers included cortical thickness in the bilateral fusiform, inferior temporal, and middle temporal gyri; thickness in the right banks of the superior temporal sulcus, the entorhinal, and the inferior parietal gyri; and thickness of the left isthmus cingulate gyrus. In addition, volumes of the left amygdala and bilateral putamen were also reduced in AD and PD cases. ROI measurements associated with AD and PD with effect sizes in opposing direction included: bilateral thickness of postcentral gyrus ,right thickness of lingual and superiofrontal gyrus, and left thickness of precentral gyrus. Of note, the two studies used slightly different covariates, Couvy-Duchesne et al,. controlled for ICV, total thickness and surface area in all models, while Laansma only controlled for ICV when testing subcortical volumes, and did not control for global thickness and surface when testing cortical ROIs.

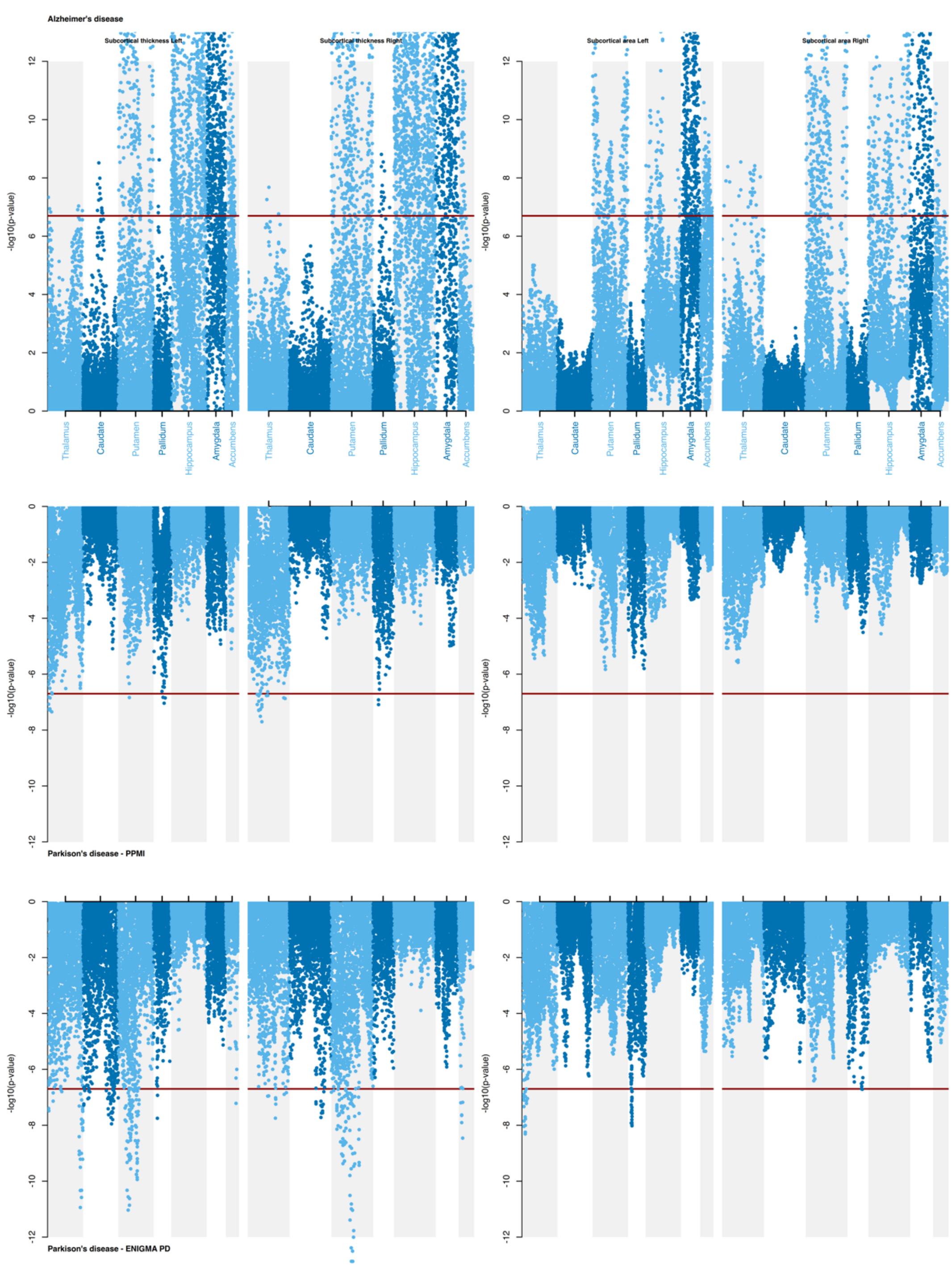

**SFigure 2: Miami plot of the subcortical vertex-wise associations for AD (top), PD (in PPMI, middle) and PD (from ENIGMA-PD, bottom).**

Associations from ENIGMA-PD included a larger sample size across multiple cohorts, leading to more vertices reaching significance than in the analysis performed in PPMI only.

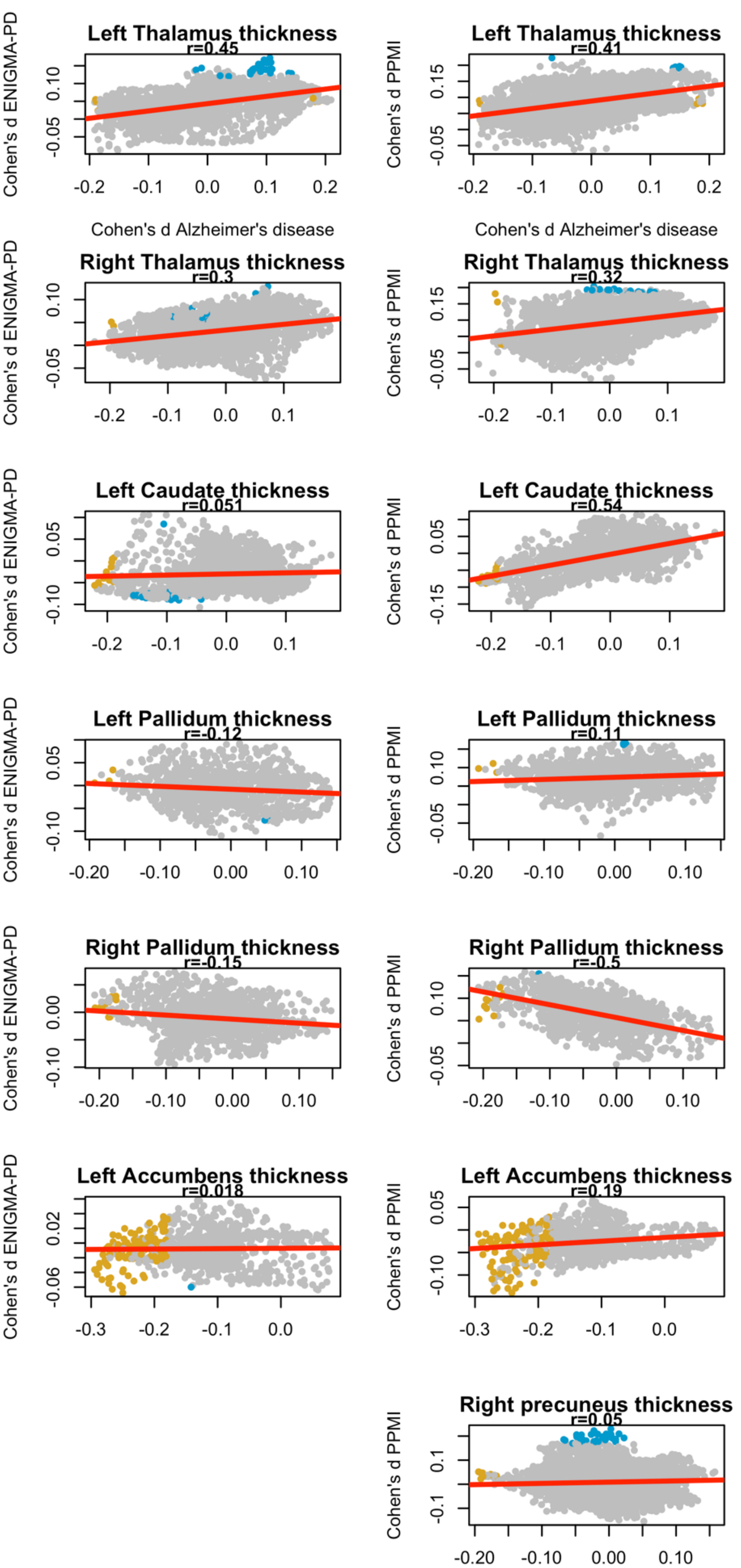

**SFigure 3: Vertex-wise association effect sizes (Cohen's d) with Alzheimer's (X-axes) and Parkinson's disease (Y-axes), in ROIs exhibiting significant but non-overlapping vertex-wise clusters for both diseases.** ROI name is listed as the panel title. Significant vertices after multiple testing correction (p-value < 2.0E-7) are highlighted in yellow (vertices significantly associated with Alzheimer's disease) and purple (vertices significantly associated with Parkinson's disease; PPMI). The first column shows PD effect sizes from ENIGMA-PD (subcortical structures only) while the second column shows the PD effect sizes from PPMI. Pearson's correlation between effect sizes is shown at the top of the plot, and the regression line is displayed in red

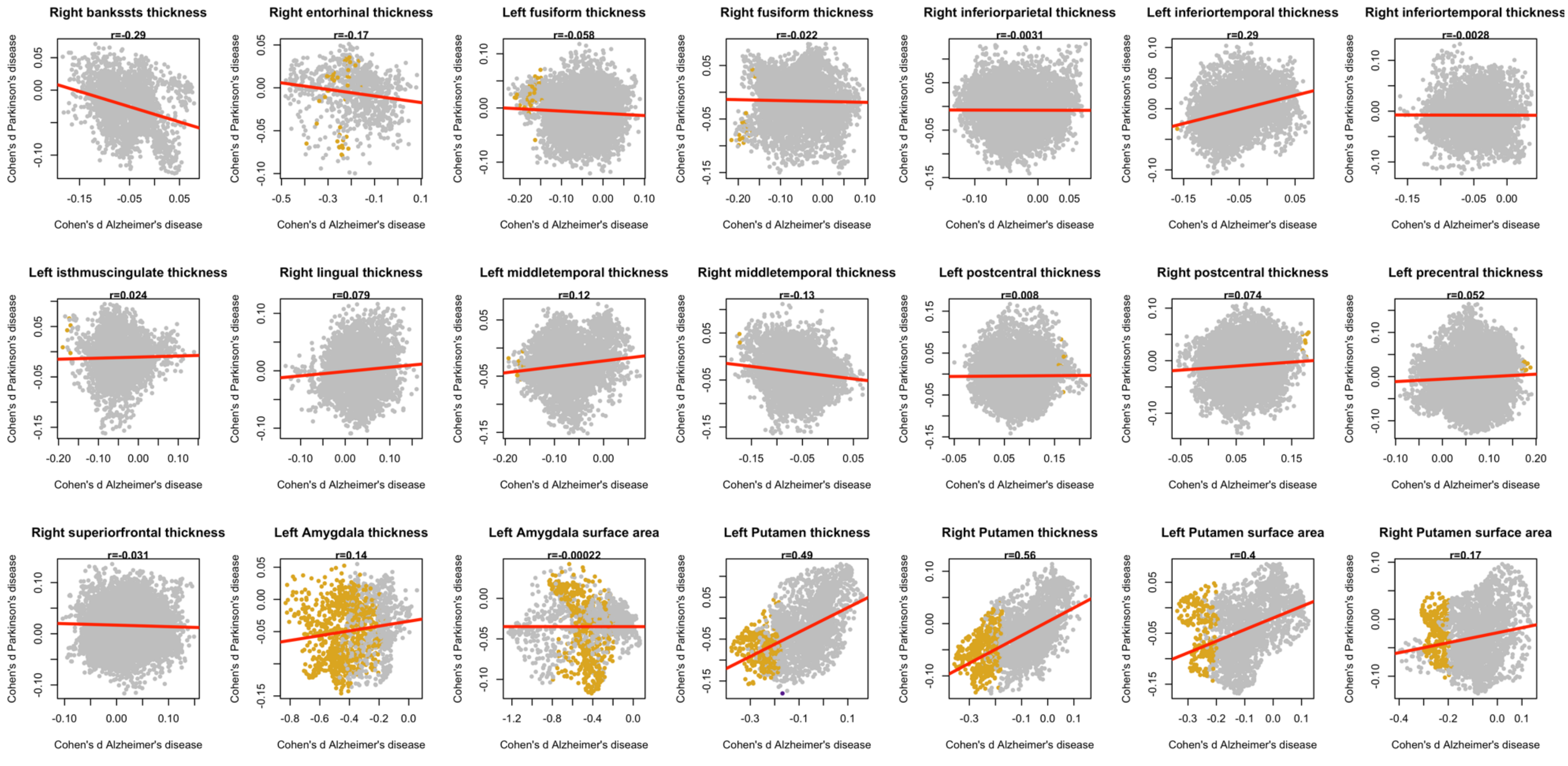

**SFigure 4. Vertex-wise association effect sizes (Cohen's d) with Alzheimer's (X-axes) and Parkinson's disease (Y-axes), in ROIs associated with AD and PD (from Laansma et al., 2021 and Couvy-Duchesne et al., 2025).**

ROI name is listed as the panel title. Significant vertices after multiple testing correction (p-value < 2.0E-7) are highlighted in yellow (vertices significantly associated with Alzheimer's disease) and purple (vertices significantly associated with Parkinson's disease; PPMI). Pearson's correlation between effect sizes is shown at the top of the plot, and the regression line is displayed in red